\documentclass[nohyper]{JHEP3}
\usepackage{cite,epsfig,amssymb,amsmath}

\newcommand{\mycomm}[1]{\hfill\break $\phantom{a}$\kern-3.5em{\tt===$>$ \bf #1}\hfill\break}
\newcommand{\mycommA}[1]{\hfill\break $\phantom{a}$\kern-3.5em{\tt   $>$ \bf #1}\hfill\break}

\newcommand{\be}{\begin{equation}}
\newcommand{\ee}{\end{equation}}
\newcommand{\ba}{\begin{eqnarray}}
\newcommand{\ea}{\end{eqnarray}}

\def\DY{\hbox{\tiny DY}}

\def\Mink{\hbox{\tiny Mink}}

\catcode`\@=11 
\def\lsim{\mathrel{\mathpalette\@versim<}}
\def\gsim{\mathrel{\mathpalette\@versim>}}
\def\@versim#1#2{\vcenter{\offinterlineskip
        \ialign{$\m@th#1\hfil##\hfil$\crcr#2\crcr\sim\crcr } }}
\catcode`\@=12 

\title{Threshold resummation to any order in $(1-x)$}

\author{Georges Grunberg\\
        Centre de Physique Th\'eorique,  \'Ecole
Polytechnique, CNRS,\\
        91128 Palaiseau Cedex, France\\
        E-mail: \email{georges.grunberg@pascal.cpht.polytechnique.fr}}

\abstract{A simple ansatz is suggested for the structure of threshold resummation of the momentum space physical evolution kernels (`physical anomalous dimensions') at all orders in $(1-x)$, taking as examples Deep Inelastic Scattering  and the Drell-Yan process. Each term in the expansion is associated to a   distinct  renormalization group and scheme invariant perturbative object (`physical Sudakov anomalous dimension') depending on a single momentum scale variable. Both logarithmically enhanced terms and constant terms are captured by the ansatz at any order in the expansion.
The ansatz is motivated by a large--$\beta_0$ dispersive calculation. A dispersive representation at finite $\beta_0$ of the physical Sudakov anomalous dimensions is also obtained, associated to a set of `Sudakov effective charges'  which encapsulate the non-Abelian nature of the interaction.  It is found that the dispersive representation requires a non-trivial, and process-dependent, choice of variables in the $(x,Q^2)$ plane.  Some interesting properties of the physical Sudakov anomalous dimensions are pointed out.
The ensuing $1/N$ expansion in moment space is straightforwardly derived from the momentum space expansion. }


\preprint{CPHT-RR 157.1007}

\begin{document}

\section{Introduction}
Threshold resummation, namely the resummation to all orders of perturbation theory of the large logarithmic
corrections which arise from the incomplete cancellation of  soft and collinear gluons at the edge of phase
space, is by now a well developed topic \cite{Sterman:1986aj,Catani:1989ne} in perturbative QCD.   About ten years ago, the subject was extended \cite{Akhoury:1998gs,Sotiropoulos:1999hy,Akhoury:2003fw} to cover also the resummation of logarithmically enhanced terms which are  suppressed by some power of $(1-x)$ for $x\rightarrow 1$ in momentum space (or by some power of $1/N$, $N\rightarrow\infty$ in moment space), concentrating on the case of the longitudinal structure function $F_{L}(x,Q^2)$ in Deep Inelastic Scattering (DIS) where these corrections are actually the leading terms. As far as I am aware, little work has been performed on this subject since then. In this paper, building upon the recent work \cite{Gardi:2007ma},  I provide a very simple ansatz for the  structure of threshold resummation at all orders in $(1-x)$, working at the level of the {\em momentum space} physical evolution kernels (or `physical anomalous dimensions', see e.g. \cite{Grunberg:1982fw,DMW,Catani:1996sc,van_Neerven:1999ca}), which are infrared and collinear safe quantities describing the physical scaling violation, where the structure of the ansatz appears to be particularly transparent.  I shall deal explicitly with the examples of the (non-singlet) Deep Inelastic Scattering (DIS) structure function $F_2(x,Q^2)$,  as well as with the Drell-Yan process. The ansatz is motivated by a large--$\beta_0$ dispersive calculation, and, following \cite{Grunberg:2006hg,Grunberg:2006jx,Grunberg:2006gd,Grunberg:2006ky,Friot:2007fd,Gardi:2007ma}, easily generalizes itself to a general finite--$\beta_0$ dispersive representation of the physical evolution kernel at any order in the  $x\rightarrow 1$ expansion.

The paper is organized as follows:  section~\ref{sec:DIS} is devoted to DIS.  In section~\ref{sec:expansion-DIS},  the momentum space ansatz for threshold resummation for the non-singlet structure function $F_2(x,Q^2)$ is displayed, which introduces at each order of the $x\rightarrow 1$ expansion a new `jet  physical anomalous dimension'. The ansatz is justified in section~\ref{sec:disp-largeb-DIS} by a large--$\beta_0$ calculation, which also provides a dispersive representation for each of the previous  physical anomalous dimensions. These dispersive representations are then extended to finite--$\beta_0$ in section~\ref{sec:disp-DIS}, where a set of `jet Sudakov effective charges' which encapsulate the non-Abelian nature of the interaction is introduced. The Drell-Yan case is addressed in section~\ref{sec:DY} in quite a similar way. The ansatz for the analogous $\tau\rightarrow 1$ expansion is given in section~\ref{sec:expansion-DY}, and justified in section~\ref{sec:disp-largeb-DY} by a large--$\beta_0$ calculation, which yields the dispersive representation of the corresponding `soft  physical anomalous dimensions' specific to the Drell-Yan process. These representations are then extended to finite--$\beta_0$ in section~\ref{sec:disp-DY}, where a set of `soft  Sudakov effective charges'  is introduced. Both in the DIS and in the Drell-Yan case, a non-trivial (process-dependent) choice of expansion parameter in the $(x, Q^2)$ (resp. $(\tau, Q^2)$) plane  has to be made in order to derive the dispersive representations.
The conclusions are given in Section 4.  The $1/N$ expansion in moment space  is derived in a straighforward way  from the corresponding momentum space expansion ansatz in Appendix A (DIS) and B (Drell-Yan). These two last appendices also clarify the connection between the definitions of the `jet' (DIS) (or  the `soft'  (DY)) scales in momentum and moment spaces respectively.

\section{Deep Inelastic Scattering  case~\label{sec:DIS}}

\subsection{A systematic expansion for $x\rightarrow 1$~\label{sec:expansion-DIS}}

The scale--dependence of the (flavour non-singlet) deep inelastic structure function $F_2$ can be expressed in terms of $F_2$ itself, yielding the following evolution equation
(see e.g.~Refs.~\cite{Grunberg:1982fw,Catani:1996sc,van_Neerven:1999ca}):
\begin{equation}
\label{F_2_evolution}
\frac{dF_2(x,Q^2)}{d\ln Q^2}\,=\,\int_x^1 \frac{dz}{z}\,K(x/z,Q^2)\,F_2(z,Q^2)\ .
\end{equation}
$K(x,Q^2)$ is the momentum space \emph{physical evolution kernel}, or {\em physical anomalous dimension}; it is renormalization--group invariant. In \cite{Gardi:2007ma}, using known results of Sudakov resummation in moment space, the result for the leading contribution to this quantity in the $x\rightarrow 1$ limit was derived. For completness, I reproduce this short derivation here.

Defining moments by
\begin{equation}
\widetilde{F}_2(N,Q^2)=\int_{0}^{1} dx\, x^{N-1} F_2(x,Q^2)\,,
\end{equation}
Eq.~(\ref{F_2_evolution}) implies that the moment--space physical evolution kernel is:
\begin{equation}
\label{DIS_kernel_def}
\widetilde{K}(N,Q^2) \equiv \int_{0}^{1} dx\,  x^{N-1} K(x,Q^2)= \frac{d\ln \widetilde{F}_2(N,Q^2)}{d\ln Q^2}.
\end{equation}

Let us now consider the $N\to \infty$ limit (corresponding to $x\to 1$ in momentum space). In this limit the evolution of the structure function takes a simple from~\cite{Forte:2002ni,Gardi:2002xm,Grunberg:2006hg,Grunberg:2006jx} (a straightforward derivation from the standard Sudakov resummation formulas can be found in \cite{Friot:2007fd}):
\begin{eqnarray}
\label{DIS_physical}
 \!\!\!\!\!\!\frac{d\ln \widetilde{F}_2(N,Q^2)}{d\ln Q^2}
&=&  \int_0^1\! dx\frac{ x^{N-1}-1}{1-x} {\cal
J}\left((1-x)Q^2\right)+ H(\alpha_s( Q^2))+{\cal O}(1/N)\,
\end{eqnarray}
where the first term, which includes the $N\to \infty$ divergent corrections to all orders, is controlled by the `jet' {\em physical} Sudakov anomalous dimension ${\cal J}(\mu^2)$ (a renormalization scheme invariant quantity):
\begin{equation}\label{Sud_anom_dim_J}
\ {\cal J}(\mu^2)={\cal A}(\alpha_s(\mu^2))+\frac{d{\cal B}(\alpha_s(\mu^2))}{d\ln \mu^2}\,\end{equation}
where ${\cal A}(\alpha_s)$ is the `cusp' anomalous dimension, and ${\cal B}(\alpha_s)$ the  standard `jet'  Sudakov anomalous dimension\footnote{${\cal A}$ and ${\cal B}$ are separately scheme dependent quantities.}.
 Moreover, the constant term can be written~\cite{Friot:2007fd} in terms of the quark electromagnetic form factor ${\cal F}(Q^2)$~\cite{Sen:1981sd,Mueller:1979ih,Collins:1980ih,Magnea:1990zb}:
\begin{align}
\label{DIS_constant}
\begin{split}
H(\alpha_s( Q^2))&=\frac{d\ln \left({\cal F}(Q^2)\right)^2}{d\ln Q^2}\,+\,\int_0^{Q^2}\frac{d\mu^2}{\mu^2}{\cal J}(\mu^2)\\
&={\mathbb G}\left(1,\alpha_s(Q^2),\varepsilon=0\right)\,+\,{\cal B}\left(\alpha_s(Q^2)\right)\,,
\end{split}
\end{align}
where each of the two terms in the first line is separately infrared divergent, but the divergence cancels~\cite{Friot:2007fd} in the sum; in the second line the result is expressed in terms of ${\cal B}(\alpha_s)$  and  ${\mathbb G}\left(Q^2/\mu^2,\alpha_s(\mu^2),\varepsilon\right)$, which is the finite part of ${d\ln \left({\cal F}(Q^2)\right)^2}/{d\ln Q^2}$ as defined in Ref.~\cite{Magnea:1990zb} using dimensional regularization.

Comparing (\ref{DIS_kernel_def}) and (\ref{DIS_physical}) one therefore finds the following relation \cite{Gardi:2007ma} in momentum space:
\begin{align}
\label{K_x}
\begin{split}
K(x,Q^2)&= \frac{{\cal J}\left((1-x)Q^2\right)}{1-x}\,+\,\frac{d\ln \left({\cal F}(Q^2)\right)^2}{d\ln Q^2}\,\delta(1-x)+{\cal O}\left((1-x)^{0}\right)\\
&= \left[\frac{{\cal J}
\left((1-x)Q^2\right)}{1-x}\right]_{+} + \left(\underbrace{
\frac{d\ln \left({\cal F}(Q^2)\right)^2}{d\ln Q^2}\,+\,\int_0^{Q^2}\frac{d\mu^2}{\mu^2}{\cal J}(\mu^2)
}_{\text{infrared \,\,finite}}\right)\,\delta(1-x)\\&\hspace*{50pt} +\,{\cal O}\left((1-x)^{0}\right)\,,
\end{split}
\end{align}
where 
the integration prescription $[\,]_{+}$ is defined by
\begin{equation}
\label{plus}
\int_0^1 dx\, F(x) \left[\frac{{\cal J}
\left((1-x)Q^2\right)}{1-x}\right]_{+}=\int_0^1dx\,
\Big(F(x)-F(1)\Big)\,\left(\frac{{\cal J}
\left((1-x)Q^2\right)}{1-x}\right)\,,
\end{equation}
where $F(x)$ is a smooth test function.
This prescription accounts for the divergent virtual corrections, which cancel against the singularity generated when integrating the real--emission contributions near $x=1$.
One  thus finds that ${{\cal J}\left((1-x)Q^2\right)}/({1-x})$ is the leading term in the expansion of the physical momentum space evolution kernel $K(x,Q^2)$ in the \hbox{$x\to 1$} limit with  $(1-x)Q^2$ fixed. The  term  proportional to $\delta(1-x)$ is comprised of purely virtual corrections associated with the quark form factor. This term is infrared divergent, but as indicated in the second line in (\ref{K_x}), the singularity cancels exactly upon integrating over~$x$ with the divergence of the integral of ${{\cal J} \left((1-x)Q^2\right)}/({1-x})$ near $x\to 1$.

Next  I observe that eq.(\ref{K_x}) strongly  suggests the following generalization to a systematic expansion for $x\rightarrow 1$ in powers of $1-x$, or, more conveniently (for reasons to be clarified in section~\ref{sec:disp-largeb-DIS}), in powers of
\begin{equation}
\label{r-DIS}
r\equiv\frac{1-x}{x}
\end{equation}
 at {\em fixed} jet mass 
\begin{equation}
\label{W}
W^2\equiv r\,Q^2
\end{equation}
namely:
\begin{equation}
\label{r-expansion-DIS}
K(x,Q^2)=\frac{1}{r}\,{\cal J}\left(W^2\right)\,+\,\frac{d\ln \left({\cal F}(Q^2)\right)^2}{d\ln Q^2}\,\delta(1-x)+{\cal J}_{0}\left(W^2\right)\,+r\,{\cal J}_{1}\left(W^2\right)\,+{\cal O}\left(r^2\right),
\end{equation}
where (barring the virtual contribution) all coefficients ${\cal J}\left(W^2\right)$ and ${\cal J}_{i}\left(W^2\right)$  are renormalization group and scheme invariant `effective charges' \cite{Grunberg:1982fw}, the physical `jet' Sudakov anomalous dimensions, functions of a {\em single} variable--the jet mass $W^2$, that can be computed order by order in $\alpha_s(W^2)$. Eq.(\ref{r-expansion-DIS}) represents a very simple momentum-space version of Sudakov resummation  for the physical evolution kernel. I note that each power of $r$ could be a priori multiplied by some powers of $\ln r$. However, evidence from a large--$\beta_0$ calculation (see below) suggests this is actually not the case, and only powers of $r$ eventually do appear. According to the ansatz (\ref{r-expansion-DIS}), the whole towers of Sudakov logarithms $\ln^p(r)/r$ in leading order, as well as those which are suppressed by a power of $r$ ($r^i\ln^j(r)$, $i\geq 0$) in subleading orders, are thus generated by expanding the ${\cal J}\left(W^2\right)$ and ${\cal J}_{i}\left(W^2\right)$ physical jet anomalous dimensions in powers of a fixed (i.e. $x$-independent) coupling, say $\alpha_s(Q^2)$, i.e. are ultimately generated by standard  renormalization group logarithms. In particular, any given term in this $r$ expansion is associated to an infinite tower of arbitrary powers of $\ln r$.
Moreover, the same ansatz captures all non-logarithmic   terms, i.e. all $\ln^0(r)/r$ terms in leading order (together with the virtual contribution), as well as all terms $r^i\ln^0(r)$ ($i\geq 0$) at  subleading orders .
The conjecture that all Sudakov logarithms in the physical evolution kernel can be absorbed by the change of argument $Q^2\rightarrow W^2$ in the running coupling was already proposed in \cite{Grunberg:1982fw}, a generalization of the suggestion in \cite{Curci:1979am,Amati:1980ch}. In the leading order of the $r$ expansion, this statement is now seen to be a consequence of the standard resummation formalism \cite{Sterman:1986aj,Catani:1989ne}.
Beyond leading order, this statement remains a conjecture, which could however be checked by matching the ansatz (\ref{r-expansion-DIS}) with existing \cite{Moch:2004pa,Vermaseren:2005qc}  fixed order calculations; moreover, this comparison would determine the perturbative expansion of the ${\cal J}\left(W^2\right)$ and ${\cal J}_{i}\left(W^2\right)$ anomalous dimensions up to ${\cal O}(\alpha_s^3)$. I also note that no integration over the Landau pole explicitly appears in this momentum space version of Sudakov resummation. We shall see however that the perturbative expansions of the subleading physical Sudakov  anomalous dimensions  ${\cal J}_{i}\left(W^2\right)$  do contain infrared renormalons (at the difference of  ${\cal J}\left(W^2\right)$!).

One can also give the generalization of eq.(\ref{K_x}), with a regularized virtual contribution:
\begin{align}
\label{r-expansion-reg-DIS}
\begin{split}
K(x,Q^2)&= \frac{1}{r}\,{\cal J}\left(W^2\right)\,+\,\frac{d\ln \left({\cal F}(Q^2)\right)^2}{d\ln Q^2}\,\delta(1-x)+{\cal J}_{0}\left(W^2\right)\,+r\,{\cal J}_{1}\left(W^2\right)\,+{\cal O}\left(r^2\right)\\
&= \left[\frac{1}{r}\,{\cal J}\left(W^2\right)\right]_{+} + \left(\underbrace{
\frac{d\ln \left({\cal F}(Q^2)\right)^2}{d\ln Q^2}\,+\,\int_0^{Q^2}\frac{d\mu^2}{\mu^2}{\cal J}(\mu^2)
}_{\text{infrared \,\,finite}}\right)\,\delta(1-x)\\&\hspace*{50pt} +\,{\cal J}_{0}\left(W^2\right)+r\,{\cal J}_{1}\left(W^2\right)\,+{\cal O}\left(r^2\right)\ ,
\end{split}
\end{align}
where 
the  integration prescription $[\,]_{+}$ is now defined by ($r=\frac{1-x}{x}$, $W^2=\frac{1-x}{x}Q^2$):
\begin{equation}
\label{plus-bis}
\int_0^1 dx F(x) \left[\frac{1}{r}\,{\cal J}\left(W^2\right)\right]_{+}=\int_0^1dx\,
\Big(F(x)\,\frac{1}{r}\,{\cal J}\left(W^2\right)-F(1)\,\frac{1}{1-x}\,{\cal J}\left((1-x)Q^2\right)\Big)\,,
\end{equation}
where $F(x)$ is a smooth test function. The proof that (\ref{r-expansion-reg-DIS}) is indeed equivalent to (\ref{K_x}) for the two leading terms is given in Appendix \ref{sec:largeN-DIS}, by considering the moments of the two expressions.

\subsection{Dispersive representation of the physical Sudakov anomalous dimensions (large $\beta_0$)~\label{sec:disp-largeb-DIS}}

The correctness of the ansatz Eq.(\ref{r-expansion-DIS}) can be checked at large--$\beta_0$, upon taking the  $x\rightarrow 1$ expansion
of the large--$\beta_0$  dispersive representation of the physical evolution kernel. This procedure will actually yield a more powerful result, namely the (large--$\beta_0$) dispersive representations of the physical Sudakov anomalous dimensions ${\cal J}$ and  ${\cal J}_{i}$. 

Let us consider the large--$\beta_0$  but \emph{arbitrary}   $x$ ($0<x<1$) dispersive representation \cite{DMW} of the physical evolution kernel. At the level of the partonic calculation,
the convolution on the r.h.s of (\ref{F_2_evolution}) becomes trivial in the large--$\beta_0$ limit since the ${\cal O}(\alpha_s)$ corrections to $F_2(z,Q^2)$ generate terms that are subleading by
powers of $1/\beta_0$. Therefore, in this limit $K(x,Q^2)$ is directly proportional to the partonic ${dF_2(x,Q^2)}/{d\ln Q^2}$, so
\begin{align}
  \label{finite-r-DIS}
  \left.K(x,Q^2)\right\vert_{\rm large \,\,\beta_0}\,
  = &\,C_F\,\int_0^{\infty}\frac{d\mu^2}{\mu^2} a_V^{Mink}(\mu^2)\,x\,  \ddot{\cal F}(\mu^2/Q^2,x),
\end{align}
where ${\cal F}(\mu^2/Q^2,x)$ is the standard notation for the characteristic function corresponding to $F_2(x,Q^2)/x$ (see Eq. (4.27) in~\cite{DMW} or (3.2) in~\cite{Dasgupta:1996hh}) (with $\dot{\cal F}\equiv -\mu^2\frac{d}{d\mu^2}$) , and $a_V^{Mink}(\mu^2)$ is the integrated time-like discontinuity of the one-loop V-scheme coupling (which corresponds to the single dressed gluon propagator, using `naive non-abelization'):
\begin{equation}
\label{eq:int_discontinuity}
\rho_V(\mu^2)=\frac{d a_V^{\Mink}(\mu^2)}{d\ln\mu^2}\,;\qquad\quad
 a_V^{\Mink}(\mu^2)\equiv -\int_{\mu^2}^{\infty} \frac{dm^2}{m^2}\rho_V(m^2),
\end{equation}
where 
\begin{equation}
\label{eq:discontinuity}
\rho_V(\mu^2)\equiv  \frac{1}{\pi}\,{\rm
    Im}\left\{\alpha_s^V(-\mu^2-i0)/\pi\right\}\,
\end{equation}
and
\begin{equation}
\label{one_loop}
\frac{\alpha_s^V(k^2)}{\pi}
= \frac{1}{\beta_0}\,\frac{1}{\ln\left(
{k^2}/{\Lambda_V^2}\right)}
\ .
\end{equation}
with
\begin{equation}
\label{Lambda_V}
\Lambda_V^2= \Lambda^2
{\rm e}^{{5}/{3}}\,,
\end{equation}
where $\Lambda^2$ is defined in the $\overline{\rm MS}$ scheme, and $\beta_0=\frac{11}{12}C_A-\frac16 N_f$.

Next, one takes the $x\rightarrow 1$ expansion under the integral (\ref{finite-r-DIS}) with a fixed invariant jet mass $W^2=Q^2(1-x)/x\equiv Q^2\,r$. To achieve this, one splits the characteristic function into its real and virtual contributions:
\begin{equation}\label{char}{\cal F}(\epsilon,x)={\cal F}^{(r)}(\epsilon,x)\theta(1-x-\epsilon\,x)+{\cal V}_s(\epsilon)\,\delta(1-x)
\end{equation}
where $\epsilon\equiv \mu^2/Q^2$, and expand the real contribution  (the virtual contribution should be left unexpanded). Using the explicit expression\footnote{The present normalization of ${\cal F}(\epsilon,x)$ is half the one in \cite{DMW}.} for ${\cal F}^{(r)}(\epsilon,x)$  in \cite{DMW}, one obtains the small $r$ expansion at {\em fixed} $\xi$:
\begin{equation}
\label{r-exp-rchar}
x\,{\cal F}^{(r)}(\epsilon,x)=\frac{1}{r}\,{\cal F}^{(r)}(\xi)+{\cal F}_0^{(r)}(\xi)+r\,{\cal F}_1^{(r)}(\xi)+{\cal O}(r^2)\ ,
\end{equation}
where
\begin{equation}
\label{xi}\xi\equiv\frac{\epsilon}{r}=\frac{\mu^2}{W^2}\ ,
\end{equation}
with
\begin{align}
\label{Fr-exp}
\begin{split}
{\cal F}^{(r)}(\xi)&=-\ln \xi-\frac{3}{4}+\frac{1}{2}\,\xi+\frac{1}{4}\,\xi^2\\
{\cal F}_0^{(r)}(\xi)&=\ln \xi+\frac{7}{2}+2\,\xi\ln \xi-4\,\xi+\frac{1}{2}\,\xi^2\\
{\cal F}_1^{(r)}(\xi)&=-\frac{3}{2}\ln\xi-6-9\,\xi\ln \xi+\frac{1}{2}\,\xi-\xi^2\ln \xi+\frac{11}{2}\,\xi^2
\ .
\end{split}
\end{align}
Thus ($^.\equiv -\mu^2\frac{d}{d\mu^2}$):
\begin{equation}
\label{r-exp-drchar}
x\,\dot{\cal F}^{(r)}(\epsilon,x)=\frac{1}{r}\,\dot{\cal F}^{(r)}(\xi)+\dot{\cal F}_0^{(r)}(\xi)+r\,\dot{\cal F}_1^{(r)}(\xi)+{\cal O}(r^2)\ ,
\end{equation}
with
\begin{align}
\label{dFr-exp}
\begin{split}
\dot{\cal F}^{(r)}(\xi)&=1-\frac{1}{2}\,\xi-\frac{1}{2}\,\xi^2\\
\dot{\cal F}_0^{(r)}(\xi)&=-1-2\,\xi\ln \xi+2\,\xi-\xi^2\\
\dot{\cal F}_1^{(r)}(\xi)&=\frac{3}{2}+9\,\xi\ln \xi+\frac{17}{2}\,\xi+2\,\xi^2\ln \xi-10\,\xi^2
\ ,
\end{split}
\end{align}
and, taking a second derivative: 
\begin{equation}
\label{r-exp-ddrchar}
x\,\ddot{\cal F}^{(r)}(\epsilon,x)=\frac{1}{r}\,\ddot{\cal F}^{(r)}(\xi)+\ddot{\cal F}_0^{(r)}(\xi)+r\,\ddot{\cal F}_1^{(r)}(\xi)+{\cal O}(r^2)\ ,
\end{equation}
with
\begin{align}
\label{ddFr-exp}
\begin{split}
\ddot{\cal F}^{(r)}(\xi)&=\frac{1}{2}\,\xi+\xi^2\\
\ddot{\cal F}_0^{(r)}(\xi)&=2\,\xi\ln \xi+2\,\xi^2\\
\ddot{\cal F}_1^{(r)}(\xi)&=-9\,\xi\ln \xi-\frac{35}{2}\,\xi-4\,\xi^2\ln \xi+18\,\xi^2
\ .
\end{split}
\end{align}
Using these results in (\ref{char}), and noting that
\begin{equation}\label{theta}\theta(1-x-\epsilon\,x)=\theta(\xi<1)\ ,\end{equation} 
 one gets:
\begin{equation}
\label{r-exp-char}
x\,{\cal F}(\epsilon,x)=\Big[\frac{1}{r}\,{\cal F}_{{\cal J}}(\xi)+{\cal F}_{{\cal J}_0}(\xi)+r\,{\cal F}_{{\cal J}_1}(\xi)+{\cal O}(r^2)\Big]+{\cal V}_s(\epsilon)\,\delta(1-x)\ ,
\end{equation}
with
\begin{align}
\label{F_J}
\begin{split}
{\cal F}_{{\cal J}}(\xi)&={\cal F}^{(r)}(\xi)\,\theta(\xi<1)\\
{\cal F}_{{\cal J}_i}(\xi)&={\cal F}_i^{(r)}(\xi)\,\theta(\xi<1)
\ ,
\end{split}
\end{align}

\begin{equation}
\label{r-exp-dchar}
x\,\dot{\cal F}(\epsilon,x)=\Big[\frac{1}{r}\,\dot{\cal F}_{{\cal J}}(\xi)+\dot{\cal F}_{{\cal J}_0}(\xi)+r\,\dot{\cal F}_{{\cal J}_1}(\xi)+{\cal O}(r^2)\Big]+\dot{\cal V}_s(\epsilon)\,\delta(1-x)\ ,
\end{equation}
with
\begin{align}
\label{dF_J}
\begin{split}
\dot{\cal F}_{{\cal J}}(\xi)&=\dot{\cal F}^{(r)}(\xi)\,\theta(\xi<1)\\
\dot{\cal F}_{{\cal J}_i}(\xi)&=\dot{\cal F}_i^{(r)}(\xi)\,\theta(\xi<1)
\ ,
\end{split}
\end{align}
and

\begin{equation}
\label{r-exp-ddchar}
x\,\ddot{\cal F}(\epsilon,x)=\Big[\frac{1}{r}\,\ddot{\cal F}_{{\cal J}}(\xi)+\ddot{\cal F}_{{\cal J}_0}(\xi)+r\,\ddot{\cal F}_{{\cal J}_1}(\xi)+{\cal O}(r^2)\Big]+\ddot{\cal V}_s(\epsilon)\,\delta(1-x)\ ,
\end{equation}
with
\begin{align}
\label{ddF_J}
\begin{split}
\ddot{\cal F}_{{\cal J}}(\xi)&=\ddot{\cal F}^{(r)}(\xi)\,\theta(\xi<1)\\
\ddot{\cal F}_{{\cal J}_i}(\xi)&=\ddot{\cal F}_i^{(r)}(\xi)\,\theta(\xi<1)
\ ,
\end{split}
\end{align}
where I used the fact that all the  terms ${\cal F}^{(r)}(\xi)$, ${\cal F}_i^{(r)}(\xi)$ and $\dot{\cal F}^{(r)}(\xi)$, $\dot{\cal F}_i^{(r)}(\xi)$ in (\ref{r-exp-rchar}) and (\ref{r-exp-drchar})  vanish at $\xi=1$ (which allows to treat the $\theta$ function effectively as a multiplicative constant when taking the derivatives). These features actually follow from the stronger property that the {\em exact}  function ${\cal F}^{(r)}(\epsilon,x)$, as well as its first derivative $\dot{\cal F}^{(r)}(\epsilon,x)$, {\em both vanish} at $\xi=1$, i.e. for $\epsilon=\frac{1-x}{x}$. Indeed one finds, expanding at fixed $r$:
\begin{equation}\label{xi-exp-rchar}x\,{\cal F}^{(r)}(\epsilon,x)=\frac{1}{4\,r}\,(1+r)\,(3+r)\,(1-\xi)^2+{\cal O}\left((1-\xi)^3\right)\ .\end {equation}
I further note that the terms which vanish for $\xi\rightarrow 0$ in  ${\cal F}_{{\cal J}_i}$ (but not in ${\cal F}_{{\cal J}}$!) are {\em logarithmically enhanced}, hence non-analytic, which implies (see below) that the subleading  physical Sudakov anomalous dimensions ${\cal J}_i$  {\em do have renormalons} (at the difference of the leading physical anomalous dimension ${\cal J}$!).

Reporting the result Eq. (\ref{r-exp-ddchar}) into (\ref{finite-r-DIS}), one  thus obtains the small $r$ expansion ($r=(1-x)/x$) of the physical evolution kernel:

\begin{align}
 \label{r-exp-ker}
 \begin{split}
 \left. K(x,Q^2)\right\vert_{\rm large \,\,\beta_0}\,
   &=\,C_F\,\Big\{\frac{1}{r}\int_0^{\infty}\frac{d\mu^2}{\mu^2}a_V^{Mink}(\mu^2)\,  \ddot{\cal F}_{{\cal J}}(\xi)+\int_0^{\infty}\frac{d\mu^2}{\mu^2}a_V^{Mink}(\mu^2)\,  \ddot{\cal F}_{{\cal J}_0}(\xi)\\
   &+r\int_0^{\infty}\frac{d\mu^2}{\mu^2}a_V^{Mink}(\mu^2)\,  \ddot{\cal F}_{{\cal J}_1}(\xi)+{\cal O}(r^2)\Big\}\\
   &+\delta(1-x)\,C_F\int_0^{\infty}\frac{d\mu^2}{\mu^2}a_V^{Mink}(\mu^2) \ddot {\cal V}_s(\epsilon)\ .
\end{split}
\end{align}
Comparing with (\ref{r-expansion-DIS}) I deduce ($\xi=\mu^2/W^2$):
\begin{align}
\label{disp-rep-J}
 \begin{split}
 \left. {\cal J}(W^2)\right\vert_{\rm large \,\,\beta_0}\,
 &=\,C_F\int_0^{\infty}\frac{d\mu^2}{\mu^2}a_V^{Mink}(\mu^2)  \,  \ddot{\cal F}_{{\cal J}}(\xi)\\
\left. {\cal J}_{i}(W^2)\right\vert_{\rm large \,\,\beta_0}\,
&=\,C_F\int_0^{\infty}\frac{d\mu^2}{\mu^2}a_V^{Mink}(\mu^2) \,  \ddot{\cal F}_{{\cal J}_i}(\xi)
\ ,
\end{split}
\end{align}
and also   ($\epsilon=\mu^2/Q^2$):
\begin{equation}\label{disp-ff}
\left. \frac{d\ln \left({\cal F}(Q^2)\right)^2}{d\ln Q^2}\right\vert_{\rm large \,\,\beta_0}\,=C_F\int_0^{\infty}\frac{d\mu^2}{\mu^2}a_V^{Mink}(\mu^2) \ddot {\cal V}_s(\epsilon)\ ,
\end{equation}
which checks Eq.(\ref{r-expansion-DIS}) in the large--$\beta_0$ limit, and in addition  gives the (large--$\beta_0$) dispersive representations of the physical Sudakov anomalous dimensions ${\cal J}$ and ${\cal J}_{i}$, as well as the (large--$\beta_0$) dispersive representation  of the quark form factor \cite{Friot:2007fd}. I note that the integral on the r.h.s. of (\ref{disp-ff}) is ultraviolet convergent, but infrared divergent, as expected from the general properties~\cite{Sen:1981sd,Mueller:1979ih,Collins:1980ih,Magnea:1990zb} of the quark form factor. Indeed from the explicit expression in \cite{DMW} one gets: $\ddot {\cal V}_s(\epsilon)={\cal O}(\ln\epsilon/\epsilon)$ for $\epsilon\rightarrow\infty$, whereas
$\ddot {\cal V}_s(\epsilon)\rightarrow -1$ for $\epsilon\rightarrow 0$. For completness, I also give the dispersive representation of the regularized virtual contribution of Eq.~(\ref{r-expansion-reg-DIS}). After a little algebra one finds, using (\ref{disp-rep-J}) and (\ref{disp-ff}):
\begin{equation}\label{disp-virt-reg}\frac{d\ln \left({\cal F}(Q^2)\right)^2}{d\ln Q^2}\,+\,\int_0^{Q^2}\frac{d\mu^2}{\mu^2}{\cal J}(\mu^2)= C_F\int_0^{\infty}\frac{d\mu^2}{\mu^2}a_V^{Mink}(\mu^2)\left( \ddot {\cal V}_s(\mu^2/Q^2)+\dot{\cal F}_{{\cal J}}(\mu^2/Q^2)\right)
\ ,
\end{equation}
where the integral on the r.h.s. is indeed both infrared and ultraviolet convergent.

\subsection{Dispersive representation of the physical Sudakov anomalous dimensions (finite $\beta_0$)~\label{sec:disp-DIS}}
Following \cite{Gardi:2007ma, Grunberg:2006gd}, it is straightforward to give the generalization of Eqs.~(\ref{disp-rep-J}) and (\ref{disp-ff}) at finite $\beta_0$:

\begin{align}
\label{disp-rep-J-finiteb}
 \begin{split}
 {\cal J}(W^2)\,
 &=\,C_F\int_0^{\infty}\frac{d\mu^2}{\mu^2}a_{{\cal J}}^{Mink}(\mu^2)  \,  \ddot{\cal F}_{{\cal J}}(\xi)\\
 {\cal J}_{i}(W^2)\,
&=\,C_F\int_0^{\infty}\frac{d\mu^2}{\mu^2}a_{{\cal J}_{i}}^{Mink}(\mu^2)\,  \ddot{\cal F}_{{\cal J}_i}(\xi)\ ,
\end{split}
\end{align}
where, in full analogy with (\ref{eq:int_discontinuity}),
\begin{equation}
\label{eq:int_discontinuity_JJi}
\rho_{{\cal J},\,{\cal J}_i}(\mu^2)=\frac{d a_{{\cal J},\,{\cal J}_i}^{\Mink}(\mu^2)}{d\ln\mu^2}\,;\qquad\quad
 a_{{\cal J},\,{\cal J}_i}^{\Mink}(\mu^2)\equiv -\int_{\mu^2}^{\infty} \frac{dm^2}{m^2}\rho_{{\cal J},\,{\cal J}_i}(m^2)\,,
\end{equation}
and, similarly to (\ref{eq:discontinuity}),  $\rho_{{\cal J},\,{\cal J}_i}(\mu^2)$ correspond to the timelike discontinuities of some ``Euclidean'' effective charges, originally defined for spacelike momenta:
\begin{align}
  \label{Eucl-Mink-J}
  a_{{\cal J},\,{\cal J}_i}^{Eucl}( k^2)\,
  &= \int_0^{\infty}\frac{d\mu^2}{\mu^2} \,a_{{\cal J},\,{\cal J}_i}^{\Mink}(\mu^2) \, \frac{\mu^2/k^2}{(1+\mu^2/k^2)^2}
  = -\int_0^{\infty}\frac{d\mu^2}{\mu^2+k^2} \,\rho_{{\cal J},\,{\cal J}_i}(\mu^2)\,.
\end{align}
Furthermore, the generalization of  (\ref{disp-ff}) is
($\epsilon=\mu^2/Q^2$):
\begin{equation}\label{disp-ff-finiteb}
 \frac{d\ln \left({\cal F}(Q^2)\right)^2}{d\ln Q^2}\,=C_F\int_0^{\infty}\frac{d\mu^2}{\mu^2}a_{{\cal F}}^{Mink}(\mu^2) \ddot {\cal V}_s(\epsilon)\ .
\end{equation}
The generalization of (\ref{disp-virt-reg}) is:
\begin{align}
\label{disp-virt-reg-finiteb}
\begin{split}
\frac{d\ln \left({\cal F}(Q^2)\right)^2}{d\ln Q^2}\,+\,\int_0^{Q^2}\frac{d\mu^2}{\mu^2}{\cal J}(\mu^2)=&C_F\int_0^{\infty}\frac{d\mu^2}{\mu^2}\Big[a_{\cal F}^{Mink}(\mu^2)\left( \ddot {\cal V}_s(\mu^2/Q^2)+1\right)\\
&+a_{\cal J}^{Mink}(\mu^2)\left(\dot{\cal F}_{{\cal J}}(\mu^2/Q^2)-1\right)\Big]
\ ,
\end{split}
\end{align}
where terms have been arranged properly to have both infrared and ultraviolet convergence: I note that the integrals over $a_{\cal F}^{Mink}(\mu^2)$ and $a_{\cal J}^{Mink}(\mu^2)$ in (\ref{disp-virt-reg-finiteb}), although infrared convergent, are separately ultraviolet divergent.

\noindent From these dispersive representations,   the perturbative expansions of the `jet'  Sudakov effective charges 
$a_{\cal J}^{Mink}(\mu^2)$ and $a_{{\cal J}_i}^{Mink}(\mu^2)$ can be derived in a straightforward way \cite{Gardi:2007ma} order by order in the full non-Abelian theory, given the expansions of ${\cal J}(W^2)$ and ${\cal J}_i(W^2)$ (and similarly for $a_{\cal F}^{Mink}(\mu^2)$ using (\ref{disp-virt-reg-finiteb}) and (\ref{DIS_constant})).

\section{Drell-Yan case~\label{sec:DY}}

\subsection{The $\tau\rightarrow 1$ expansion~\label{sec:expansion-DY}}

The cross section of the Drell--Yan process, $h_a+h_b\to e^+e^- +X$, is:
\begin{equation}
\label{DY}
\frac{d\sigma}{dQ^2}= \frac{4\pi\alpha_{\rm em}^2}{9Q^2s}\,
\sum_{i,j}\int_0^1 \frac{dx_i}{x_i} \frac{dx_j}{x_j}
\,f_{i/h_a}(x_i,\mu_F)\, f_{j/h_b}(x_j,\mu_F) \,g_{ij}(\tau,Q^2,\mu_F^2)
\end{equation}
where $s$ is the hadronic center--of--mass energy,
$\hat{s}=x_ix_js$ is the partonic one,
$Q^2$ is the squared mass of the lepton pair and $\tau=Q^2/\hat{s}$. The partonic threshold $\tau\to 1$ is characterized by Sudakov logarithms.

Let us define the Mellin transform of the quark--antiquark partonic cross section in (\ref{DY}),
e.g. in electromagnetic annihilation
$g_{q\bar{q}}(\tau,Q^2,\mu_F^2)=e_q^2  \left[\delta(1-\tau)+{\cal O}(\alpha_s)\right]$, by
\begin{align}
\label{G_N_def}
\begin{split}
&G_{q\bar{q}}(N,Q^2,\mu_F^2)\equiv \int_0^1d\tau\, \tau^{N-1}\,g_{q\bar{q}}(\tau,Q^2,\mu_F^2)\,.
\end{split}
\end{align}
In the $N\rightarrow\infty$ limit one can derive (see e.g.~\cite{Friot:2007fd}) the analogue of Eq.~(\ref{DIS_physical}):

\begin{align}
\begin{split}
\label{dG_N}
\frac{d\ln G_{q\bar{q}}(N,Q^2,\mu_F^2)}{d\ln Q^2}&\equiv \widetilde{K}_{\DY}(N,Q^2)\equiv
\int_{0}^{1} {d\tau} \tau^{N-1} K_{\DY}(\tau,Q^2)\\&
=
\int_0^1d\tau \frac{\tau^{N-1}-1}{1-\tau} \,2\, {\cal S}\left(Q^2(1-\tau)^2\right)+H_{\DY}(\alpha_s(Q^2))+{\cal O}(1/N)\ ,
\end{split}
\end{align}
where $K_{\DY}(\tau, Q^2)$ is the momentum space physical Drell--Yan evolution kernel defined for arbitrary $\tau$, and the $N$-dependent terms are controlled by the `soft' Sudakov \emph{physical} anomalous dimension ${\cal S}$:
\begin{equation}\label{Sud_anom_dim_S}
\ {\cal S}(\mu^2)={\cal A}(\alpha_s(\mu^2))+\frac{1}{2}\frac{d{\cal D}(\alpha_s(\mu^2))}{d\ln \mu^2}\ ,\end{equation}
where ${\cal D}$ is the standard (scheme-dependent) `soft' Sudakov anomalous dimension relevant to the Drell-Yan process.
Moreover, the constant term can be expressed in terms of the analytically--continued electromagnetic quark form factor~\cite{Friot:2007fd}:
\begin{equation}
\label{DY_constant}
 H_{\DY}(\alpha_s(Q^2))=
\frac{d\ln \left|\left({\cal F}(-Q^2)\right)\right|^2}{d\ln Q^2}\,+\int_0^{Q^2}\frac{d\mu^2}{\mu^2}{\cal S}(\mu^2)\,,
\end{equation}
where the infrared singularities cancel~\cite{Friot:2007fd} in the sum, as in (\ref{DIS_constant}):
\begin{align}
\label{DY_constant}
\begin{split}
H_{\DY}(\alpha_s( Q^2))&=\frac{d\ln \left|\left({\cal F}(-Q^2)\right)\right|^2}{d\ln Q^2}\,+\int_0^{Q^2}\frac{d\mu^2}{\mu^2}{\cal S}(\mu^2)\\
&=\left[{\mathbb G}\left(1,\alpha_s(Q^2),\varepsilon=0\right)\,+\beta(\alpha_s)\frac{d{\cal R}}{d\alpha_s}(\alpha_s(Q^2))\right]+\frac{1}{2}\,{\cal D}\left(\alpha_s(Q^2)\right)\, ,
\end{split}
\end{align}
with ${\cal R}(\alpha_s(Q^2))=\ln\left\vert\frac{{\cal F}(-Q^2)}{{\cal F}(Q^2)}\right\vert^2$.
Thus, in momentum space we have \cite{Gardi:2007ma}:
\begin{align}
\label{K_x_DY}
\begin{split}
 K_{\DY}(\tau,Q^2)&=
2\,\frac{
{\cal S}\left(Q^2(1-\tau)^2\right)}{1-\tau}
+
\frac{d\ln \left|\left({\cal F}(-Q^2)\right)\right|^2}{d\ln Q^2}\,\delta(1-\tau)+{\cal O}\left((1-\tau)^0\right)
\\
&=
2\,\left[\frac{
{\cal S}\left(Q^2(1-\tau)^2\right)}{1-\tau}
\right]_{+} \\&
+
\left(
\underbrace{
\frac{d\ln \left|\left({\cal F}(-Q^2)\right)\right|^2}{d\ln Q^2}\!+\!
\int_0^{Q^2}\frac{d\mu^2}{\mu^2}{\cal S}_{\DY}(\mu^2)}_{\text{infrared\,\,finite}}
\right)\,\delta(1-\tau)+{\cal O}\left((1-\tau)^0\right)
\end{split}
\end{align}
We see that the physical Sudakov anomalous dimension ${\cal S}$ controls the leading term in the expansion of the momentum space physical Drell--Yan kernel (\ref{dG_N}) near threshold. The  $\tau\to 1$
limit is taken such that $E_{\DY}=Q(1-\tau)$, corresponding to the total energy carried by soft gluons to the final state, is kept fixed.
The virtual contribution, proportional to $\delta(1-\tau)$, is determined by the quark form factor, analytically--continued to the time--like axis. This  term is infrared singular, but upon performing an integral over $\tau$ this singularity cancels with the one generated by integrating the real-emission term $2\, {\cal S}\left(Q^2(1-\tau)^2\right)/({1-\tau})$ near $\tau\to 1$.

Similarly to the DIS case, Eq.~(\ref{K_x_DY}) suggests a generalization to an expansion for $\tau\rightarrow 1$ in powers of $1-\tau$ at fixed $E_{\DY}$. 
 As in the DIS case, however, it turns out (see section~\ref{sec:disp-largeb-DY}) that in order to derive a dispersive representation such an expansion  is {\em not} the appropriate one. Instead, one should consider an expansion in powers of the alternative variable:
  \begin{equation}\label{r-tilde}r_{\DY}\equiv 2\,\frac{1-\tau}{\tau+\sqrt{\tau}}\end{equation}
  (properly normalized  so that  $r_{\DY}\sim 1-\tau$ for $\tau\rightarrow 1$)
   at {\em fixed} $W_{\DY}^2$, with:
 \begin{equation}\label{W-DY}W_{\DY}^2\equiv\ r_{\DY}^2\,Q^2\ ,\end{equation}
namely:
\begin{equation}
\label{r-tilde-expansion-DY}
K_{\DY}(\tau,Q^2)=\frac{2}{r_{\DY}}\,{\cal S}\left(W_{\DY}^2\right)\,+\,\frac{d\ln \left|\left({\cal F}(-Q^2)\right)\right|^2}{d\ln Q^2}\,\delta(1-\tau)+\,{\cal S}_{0}\left(W_{\DY}^2\right)\,+r_{\DY}\,{\cal S}_{1}\left(W_{\DY}^2\right)\,+{\cal O}\left(r_{\DY}^2\right)\ ,
\end{equation}
where the coefficients ${\cal S}$ and ${\cal S}_i$ are the physical `soft' anomalous dimensions appropriate to the Drell-Yan process. 
After regularizing the virtual contribution, Eq.~(\ref{r-tilde-expansion-DY}) can be written as:
\begin{align}
\label{r-expansion-reg-DY}
\begin{split}
K_{\DY}(\tau,Q^2)&= \frac{2}{r_{\DY}}\,{\cal S}\left(W_{\DY}^2\right)\,+\,\frac{d\ln \left|\left({\cal F}(-Q^2)\right)\right|^2}{d\ln Q^2}\,\delta(1-\tau)+\,{\cal S}_{0}\left(W_{\DY}^2\right)\,+r_{\DY}\,{\cal S}_{1}\left(W_{\DY}^2\right)\,+{\cal O}\left(r_{\DY}^2\right)\\
&= \left[\frac{2}{r_{\DY}}\,{\cal S}\left(W_{\DY}^2\right)\right]_{+} + \left(\underbrace{
\frac{d\ln \left|\left({\cal F}(-Q^2)\right)\right|^2}{d\ln Q^2}\,+\,\int_0^{Q^2}\frac{d\mu^2}{\mu^2}{\cal S}(\mu^2)
}_{\text{infrared \,\,finite}}\right)\,\delta(1-\tau)\\&\hspace*{50pt} +\,{\cal S}_{0}\left(W_{\DY}^2\right)+r_{\DY}\,{\cal S}_{1}\left(W_{\DY}^2\right)\,+{\cal O}\left(r_{\DY}^2\right)\ ,
\end{split}
\end{align}
where 
the  integration prescription $[\,]_{+}$ is  defined (similarly to (\ref{plus-bis})) by ($r_{\DY}=2\,\frac{1-\tau}{\tau+\sqrt{\tau}}$, $W_{\DY}^2=r_{\DY}^2\,Q^2$):
\begin{equation}
\label{plus-DY}
\int_0^1 d\tau F(\tau) \left[\frac{1}{r_{\DY}}\,{\cal S}\left(W_{\DY}^2\right)\right]_{+}=\int_0^1d\tau\,
\left[F(\tau)\,\frac{1}{r_{\DY}}\,{\cal S}\left(W_{\DY}^2\right)-F(1)\,\frac{1}{1-\tau}\,{\cal S}\left((1-\tau)^2Q^2\right)\right]\,,
\end{equation}
where $F(\tau)$ is a smooth test function. 

\subsection{Dispersive representation of the physical Sudakov anomalous dimensions (large $\beta_0$)~\label{sec:disp-largeb-DY}}

In quite a similar way to the DIS case, we start from the large--$\beta_0$  but \emph{arbitrary $\tau$} ($0<\tau<1$) dispersive representation of the physical Drell--Yan momentum space evolution kernel,
\begin{align}
  \label{finite-r-DY}
  \left. K_{\DY}(\tau,Q^2)\right\vert_{\rm large \,\,\beta_0}\,
  = &\,C_F\,\int_0^{\infty}\frac{d\mu^2}{\mu^2} \rho_V(\mu^2)  \left(\dot{\cal F}_{\DY}(\mu^2/Q^2,\tau)-\dot{\cal F}_{\DY}(0,\tau)\right)\ ,
\end{align}
where ${\cal F}_{\DY}(\mu^2/Q^2,\tau)$ is the characteristic function corresponding to
$g_{q\bar{q}}(\tau,Q^2,\mu_F^2)$. Here  one has to use \cite{Gardi:2007ma} this version of the dispersive representation involving the first derivative of the characteristic function, the second derivative being too singular, as we shall see, on the phase-space boundary.

Let us first take the $\tau\to 1$ expansion
 under the integral while fixing the total energy radiated into the final state, $E_{\DY}=Q(1-\tau)$.
Starting from:
\begin{equation}\label{char-DY}{\cal F}_{\DY}(\epsilon,\tau)={\cal F}_{\DY}^{(r)}(\epsilon,\tau)\,\theta\left((1-\tau(1+\sqrt{\epsilon})^2\right)+{\cal V}_t(\epsilon)\,\delta(1-\tau)\ ,
\end{equation}
where $\epsilon=\mu^2/Q^2$, and expanding the real contribution using the explicit expression\footnote{The present normalization of ${\cal F}_{\DY}$ is half the one  in \cite{DMW}.} for ${\cal F}_{\DY}^{(r)}(\epsilon,\tau)$  in \cite{DMW} (Eq.~(4.87) there), one obtains: 
\begin{align}
\label{r-exp-rchar-DY}
\begin{split}
{\cal F}_{\DY}^{(r)}(\epsilon,\tau)
=&\frac{4}{1-\tau}\textrm{tanh}^{-1}\left(\sqrt{1-\,\frac{4\,\mu^2}{E_{\DY}^2}}\,\right)\,+4\,\left(1-\frac{\mu^2}{E_{\DY}^2}\right)\left(\frac{1}{\sqrt{1-\,\frac{4\,\mu^2}{E_{\DY}^2}}}-\textrm{tanh}^{-1}\left(\sqrt{1-\,\frac{4\,\mu^2}{E_{\DY}^2}}\,\right)\right)\\
+&\,\,\,\,\,\,\,{\cal O}(1-\tau)\ .
\end{split}
\end{align}
 I note that beginning at next-to-leading order (the ${\cal O}((1-\tau)^0)$ contribution), the expansion becomes singular at $4\,\mu^2/E_{\DY}^2=1$, which corresponds to the phase-space boundary in the $\tau\rightarrow 1$ limit. Although at ${\cal O}((1-\tau)^0)$ the singularity is still\footnote{Non-integrable singularities start appearing at order ${\cal O}(1-\tau)$.} integrable, it will become non-integrable after taking one derivative will respect to $\mu^2$ and inserted into the integral on the r.h.s. of Eq.~(\ref{finite-r-DY})!

\noindent The solution to this difficulty consists in performing a change of variables. Namely, instead of $1-\tau$ and $\mu^2/E_{\DY}^2$, one should use $r_{\DY}=2\,\frac{1-\tau}{\tau+\sqrt{\tau}}$ and 
\begin{equation}\label{chi-DY}\xi_{\DY}=\frac{\epsilon}{r_{\DY}^2}=\frac{\mu^2}{W_{\DY}^2}\ ,\end{equation} 
and perform the $r_{\DY}\rightarrow 0$ expansion in powers of $r_{\DY}$, at {\em fixed} $\xi_{\DY}$. The result is:
\begin{equation}
\label{rho-exp-rchar-DY}
{\cal F}_{\DY}^{(r)}(\epsilon,\tau)=\frac{2}{r_{\DY}}\,{\cal F}_{\DY}^{(r)}(\xi_{\DY})+{\cal F}_{0,\DY}^{(r)}(\xi_{\DY})+r_{\DY}\,{\cal F}_{1,\DY}^{(r)}(\xi_{\DY})+{\cal O}(r_{\DY}^2)\ ,
\end{equation}
with
\begin{align}
\label{Fr-exp-DY}
\begin{split}
{\cal F}_{\DY}^{(r)}(\xi_{\DY})&=2\,\textrm{tanh}^{-1}(\sqrt{1-4\,\xi_{\DY}}\,)\\
{\cal F}_{0,\DY}^{(r)}(\xi_{\DY})&=\sqrt{1-4\,\xi_{\DY}}-(1-4\,\xi_{\DY})\,\textrm{tanh}^{-1}(\sqrt{1-4\,\xi_{\DY}}\,)\\
{\cal F}_{1,\DY}^{(r)}(\xi_{\DY})&=\frac{1}{8}\left(\frac{-19+84\,\xi_{\DY}-32\,\xi_{\DY}^2}{\sqrt{1-4\,\xi_{\DY}}}+(18+16\,\xi_{\DY}+32\,\xi_{\DY}^2)\,\textrm{tanh}^{-1}(\sqrt{1-4\,\xi_{\DY}}\,)\right)
\ ,
\end{split}
\end{align}
where {\em all} the terms ${\cal F}_{\DY}^{(r)}(\epsilon,\tau)$ and ${\cal F}_{i,\DY}^{(r)}(\xi_{\DY})$ now vanish at the phase-space boundary $4\,\xi_{\DY}=1$. I actually checked the stronger {\em exact}  property that ${\cal F}_{\DY}^{(r)}(\epsilon,\tau)\equiv 0$ for $4\,\xi_{\DY}=1$, which explains the previous facts. Namely one finds, expanding at fixed $r_{\DY}$ for  $4\,\xi_{\DY}\rightarrow 1$:
\begin{equation}
\label{xi-exp-rchar-DY}
{\cal F}_{\DY}^{(r)}(\epsilon,\tau)=\frac{4}{r_{\DY}}\,\left(1+\frac{1}{4}\,r_{\DY}^2\right)\,\sqrt{1+\frac{1}{2}\,r_{\DY}}\,\,\sqrt{1-4\,\xi_{\DY}}+...\ .
\end{equation}
Taking one derivative ($^.=-\mu^2\frac{d}{d\mu^2}$) one thus gets the expansion:
\begin{equation}
\label{rho-exp-drchar-DY}
\dot{\cal F}_{\DY}^{(r)}(\epsilon,\tau)=\frac{2}{r_{\DY}}\,\dot{\cal F}_{\DY}^{(r)}(\xi_{\DY})+\dot{\cal F}_{0,\DY}^{(r)}(\xi_{\DY})+r_{\DY}\,\dot{\cal F}_{1,\DY}^{(r)}(\xi_{\DY})+{\cal O}(r_{\DY}^2)\ ,
\end{equation}
with
\begin{align}
\label{dFr-exp-DY}
\begin{split}
\dot{\cal F}_{\DY}^{(r)}(\xi_{\DY})&=\frac{1}{\sqrt{1-4\,\xi_{\DY}}\,}\\
\dot{\cal F}_{0,\DY}^{(r)}(\xi_{\DY})&=\frac{2\,\xi_{\DY}}{\sqrt{1-4\,\xi_{\DY}}\,}\,-\frac{1}{2}\sqrt{1-4\,\xi_{\DY}}-\,4\,\xi_{\DY}\,\textrm{tanh}^{-1}(\sqrt{1-4\,\xi_{\DY}}\,)\\
\dot{\cal F}_{1,\DY}^{(r)}(\xi_{\DY})&=\frac{9-38\,\xi_{\DY}+64\,\xi_{\DY}^2}{8\,\sqrt{1-4\,\xi_{\DY}}}-\,2\,\xi_{\DY}(1+4\,\xi_{\DY})\,\textrm{tanh}^{-1}(\sqrt{1-4\,\xi_{\DY}}\,)
\ ,
\end{split}
\end{align}
where {\em all} the terms $\dot{\cal F}_{\DY}^{(r)}(\xi_{\DY})$ and $\dot{\cal F}_{i,\DY}^{(r)}(\xi_{\DY})$ are now singular, but integrable, at the phase-space boundary $4\,\xi_{\DY}=1$. This result follows immediately from (\ref{xi-exp-rchar-DY}) which yields:

\begin{equation}\label{xi-exp-drchar-DY}\dot{\cal F}_{\DY}^{(r)}(\epsilon,\tau)=\frac{2}{r_{\DY}}\,\left(1+\frac{1}{4}\,r_{\DY}^2\right)\,\sqrt{1+\frac{1}{2}\,r_{\DY}}\,\frac{1}{\sqrt{1-4\,\xi_{\DY}}}+...\  .
\end{equation}

Using these results in (\ref{char-DY}), and noting that:
\begin{equation}\label{theta-DY}\theta\left((1-\tau(1+\sqrt{\epsilon})^2\right)=\theta(4\,\xi_{\DY}<1)\ ,\end{equation}
one gets:
\begin{equation}
\label{r-exp-char-DY}
{\cal F}_{\DY}(\epsilon,\tau)=\Big[\frac{2}{r_{\DY}}\,{\cal F}_{{\cal S}}(\xi_{\DY})+{\cal F}_{{\cal S}_0}(\xi_{\DY})+r_{\DY}\,{\cal F}_{{\cal S}_1}(\xi_{\DY})+{\cal O}(r_{\DY}^2)\Big]+{\cal V}_t(\epsilon)\,\delta(1-\tau)\ ,
\end{equation}
with
\begin{align}
\label{F_S}
\begin{split}
{\cal F}_{{\cal S}}(\xi_{\DY})&={\cal F}_{\DY}^{(r)}(\xi_{\DY})\,\theta(4\,\xi_{\DY}<1)\\
{\cal F}_{{\cal S}_i}(\xi_{\DY})&={\cal F}_{i,\DY}^{(r)}(\xi_{\DY})\,\theta(4\,\xi_{\DY}<1)
\ ,
\end{split}
\end{align}
and

\begin{equation}
\label{r-exp-dchar-DY}
\dot{\cal F}_{\DY}(\epsilon,\tau)=\Big[\frac{2}{r_{\DY}}\,\dot{\cal F}_{{\cal S}}(\xi_{\DY})+\dot{\cal F}_{{\cal S}_0}(\xi_{\DY})+r_{\DY}\,\dot{\cal F}_{{\cal S}_1}(\xi_{\DY})+{\cal O}(r_{\DY}^2)\Big]+\dot{\cal V}_t(\epsilon)\,\delta(1-\tau)\ ,
\end{equation}
with
\begin{align}
\label{dF_S}
\begin{split}
\dot{\cal F}_{{\cal S}}(\xi_{\DY})&=\dot{\cal F}_{\DY}^{(r)}(\xi_{\DY})\,\theta(4\,\xi_{\DY}<1)\\
\dot{\cal F}_{{\cal S}_i}(\xi_{\DY})&=\dot{\cal F}_{i,\DY}^{(r)}(\xi_{\DY})\,\theta(4\,\xi_{\DY}<1)
\ ,
\end{split}
\end{align}
where I used the previously mentioned fact that the  terms  on the r.h.s. of (\ref{rho-exp-rchar-DY}) all vanish at $4\,\xi_{\DY}=1$ (which allows to treat the $\theta$ function effectively as a multiplicative constant when taking the derivatives). 
I further note that the terms which vanish for $\xi_{\DY}\rightarrow 0$ in  ${\cal F}_{{\cal S}_i}$ (but not in ${\cal F}_{{\cal S}}$!) are {\em logarithmically enhanced}, hence non-analytic, which implies (see below) that the subleading  physical Sudakov anomalous dimensions ${\cal S}_i$  {\em do have renormalons} (at the difference of the leading physical anomalous dimension ${\cal S}$!). Indeed we have for $\xi_{\DY}\rightarrow 0$:
\begin{equation}\label{log}\xi_{\DY}\,\textrm{tanh}^{-1}(\sqrt{1-4\,\xi_{\DY}}\,)\sim -\frac{1}{2}\,\xi_{\DY}\,\ln(\xi_{\DY})\ .
\end{equation}

Reporting the result (\ref{r-exp-dchar-DY}) into (\ref{finite-r-DY}), and performing the $r_{\DY}\rightarrow 0$ expansion ($r_{\DY}=2\,\frac{1-\tau}{\tau+\sqrt{\tau}}$) under the integral (which is now legitimate, since the singularities of the integrand are all integrable), one thus obtains the expansion of the physical evolution kernel:
\begin{align}
 \label{r-exp-ker-DY}
 \begin{split}
 \left. K_{\DY}(\tau,Q^2)\right\vert_{\rm large \,\,\beta_0}\,
   &=\,C_F\,\Big\{\frac{2}{r_{\DY}}\int_0^{\infty}\frac{d\mu^2}{\mu^2}\rho_V(\mu^2)\,  \left(\dot{\cal F}_{{\cal S}}(\xi_{\DY})-\dot{\cal F}_{{\cal S}}(0)\right)\\
   &+\int_0^{\infty}\frac{d\mu^2}{\mu^2}\rho_V(\mu^2)\,  \left(\dot{\cal F}_{{\cal S}_0}(\xi_{\DY})-\dot{\cal F}_{{\cal S}_0}(0)\right)\\
   &+r_{\DY}\int_0^{\infty}\frac{d\mu^2}{\mu^2}\rho_V(\mu^2)\,  \left(\dot{\cal F}_{{\cal S}_1}(\xi_{\DY})-\dot{\cal F}_{{\cal S}_1}(0)\right)+{\cal O}(r_{\DY}^2)\Big\}\\
   &+\delta(1-\tau)\,C_F\int_0^{\infty}\frac{d\mu^2}{\mu^2}a_V^{Mink}(\mu^2) \ddot {\cal V}_t(\epsilon)\ ,
\end{split}
\end{align}
where I performed integration by parts in the virtual contribution, in order to obtain at least a finite integrand, $\dot {\cal V}_t(0)$ being infinite (the integral itself is infrared divergent since $ \ddot {\cal V}_t(0)=-1$).
Comparing with (\ref{r-tilde-expansion-DY}) I deduce ($\xi_{\DY}=\mu^2/W_{\DY}^2$):
\begin{align}
\label{disp-rep-S}
 \begin{split}
 \left. {\cal S}(W_{\DY}^2)\right\vert_{\rm large \,\,\beta_0}\,
 &=\,C_F\int_0^{\infty}\frac{d\mu^2}{\mu^2}\rho_V(\mu^2)\,  \left(\dot{\cal F}_{{\cal S}}(\xi_{\DY})-\dot{\cal F}_{{\cal S}}(0)\right)\\
\left. {\cal S}_{i}(W_{\DY}^2)\right\vert_{\rm large \,\,\beta_0}\,
&=\,C_F\int_0^{\infty}\frac{d\mu^2}{\mu^2}\rho_V(\mu^2)\,  \left(\dot{\cal F}_{{\cal S}_i}(\xi_{\DY})-\dot{\cal F}_{{\cal S}_i}(0)\right)
\ ,
\end{split}
\end{align}
and also   ($\epsilon=\mu^2/Q^2$):
\begin{equation}\label{disp-ff-t}
\left. \frac{d\ln \left|\left({\cal F}(-Q^2)\right)\right|^2}{d\ln Q^2}\right\vert_{\rm large \,\,\beta_0}\,=C_F\int_0^{\infty}\frac{d\mu^2}{\mu^2}a_V^{Mink}(\mu^2) \ddot {\cal V}_t(\epsilon)\ ,
\end{equation}
which checks Eq.~(\ref{r-tilde-expansion-DY}) in the large--$\beta_0$ limit, and in addition  gives the (large--$\beta_0$) dispersive representations of the physical Sudakov anomalous dimensions ${\cal S}$ and ${\cal S}_{i}$, as well as the (large--$\beta_0$) dispersive representation  of the time-like quark form factor \cite{Friot:2007fd}. Finally, the dispersive representation of  the regularized virtual contribution in
(\ref{r-expansion-reg-DY}) is given by:
\begin{equation}\label{disp-virt-reg-t}
\frac{d\ln \left|\left({\cal F}(-Q^2)\right)\right|^2}{d\ln Q^2}\,+\,\int_0^{Q^2}\frac{d\mu^2}{\mu^2}{\cal S}(\mu^2)=C_F\int_0^{\infty}\frac{d\mu^2}{\mu^2}a_V^{Mink}(\mu^2)\left( \ddot {\cal V}_t(\mu^2/Q^2)+\dot{\cal F}_{{\cal S}}(\mu^2/Q^2)\right)
\ ,
\end{equation}
where the integral on the r.h.s. is convergent both in the infrared and the ultraviolet region.

\subsection{Dispersive representation of the physical Sudakov anomalous dimensions (finite $\beta_0$)~\label{sec:disp-DY}}
The generalization of Eq.~(\ref{disp-rep-S})  at finite $\beta_0$ is straightforward  \cite{Gardi:2007ma, Grunberg:2006gd}:
\begin{align}
\label{disp-rep-S-finiteb}
 \begin{split}
  {\cal S}(W_{\DY}^2)\,
 &=\,C_F\int_0^{\infty}\frac{d\mu^2}{\mu^2}\rho_{\cal S}(\mu^2)\,  \left(\dot{\cal F}_{{\cal S}}(\xi_{\DY})-\dot{\cal F}_{{\cal S}}(0)\right)\\
{\cal S}_{i}(W_{\DY}^2)\,
&=\,C_F\int_0^{\infty}\frac{d\mu^2}{\mu^2}\rho_{{\cal S}_i}(\mu^2)\,  \left(\dot{\cal F}_{{\cal S}_i}(\xi_{\DY})-\dot{\cal F}_{{\cal S}_i}(0)\right)
\ ,
\end{split}
\end{align}
where, in full analogy with (\ref{eq:int_discontinuity}),
\begin{equation}
\label{eq:int_discontinuity_SSi}
\rho_{{\cal S},\,{\cal S}_i}(\mu^2)=\frac{d a_{{\cal S},\,{\cal S}_i}^{\Mink}(\mu^2)}{d\ln\mu^2}\,;\qquad\quad
 a_{{\cal S},\,{\cal S}_i}^{\Mink}(\mu^2)\equiv -\int_{\mu^2}^{\infty} \frac{dm^2}{m^2}\rho_{{\cal S},\,{\cal S}_i}(m^2)\,,
\end{equation}
and, similarly to (\ref{eq:discontinuity}),  $\rho_{{\cal S},\,{\cal S}_i}(\mu^2)$ correspond to the timelike discontinuities of some ``Euclidean'' effective charges, originally defined for spacelike momenta:
\begin{align}
  \label{Eucl-Mink}
  a_{{\cal S},\,{\cal S}_i}^{Eucl}( k^2)\,
  &= \int_0^{\infty}\frac{d\mu^2}{\mu^2} \,a_{{\cal S},\,{\cal S}_i}^{\Mink}(\mu^2) \, \frac{\mu^2/k^2}{(1+\mu^2/k^2)^2}
  = -\int_0^{\infty}\frac{d\mu^2}{\mu^2+k^2} \,\rho_{{\cal S},\,{\cal S}_i}(\mu^2)\,.
\end{align}
From these dispersive representations,   the perturbative expansions of the `soft' Sudakov effective charges 
$a_{\cal S}^{Mink}(\mu^2)$ and $a_{{\cal S}_i}^{Mink}(\mu^2)$ can be derived in a straightforward way \cite{Gardi:2007ma} order by order in the full non-Abelian theory, given the expansions of ${\cal S}(W_{\DY}^2)$ and ${\cal S}_i(W_{\DY}^2)$.

Furthermore, the generalization of  (\ref{disp-ff-t}) is
\begin{equation}\label{disp-ff-t-finiteb}
 \frac{d\ln \left|\left({\cal F}(-Q^2)\right)\right|^2}{d\ln Q^2}\,=C_F\int_0^{\infty}\frac{d\mu^2}{\mu^2}a_{\cal F}^{Mink}(\mu^2) \ddot {\cal V}_t(\epsilon)\ ,
\end{equation}
whereas that of (\ref{disp-virt-reg-t}) is:
\begin{align}
\label{disp-virt-reg-t-finiteb}
\begin{split}
\frac{d\ln \left|\left({\cal F}(-Q^2)\right)\right|^2}{d\ln Q^2}\,+\,\int_0^{Q^2}\frac{d\mu^2}{\mu^2}{\cal S}(\mu^2)=&C_F\int_0^{\infty}\frac{d\mu^2}{\mu^2}\Big[a_{\cal F}^{Mink}(\mu^2)\left( \ddot {\cal V}_t(\mu^2/Q^2)+1\right)\\
&+a_{\cal S}^{Mink}(\mu^2)\left(\dot{\cal F}_{{\cal S}}(\mu^2/Q^2)-1\right)\Big]
\ ,
\end{split}
\end{align}
where terms have been arranged properly to have both infrared and ultraviolet convergence.

\section{Conclusions}
I presented a simple {\em momentum space} ansatz  for the structure of a systematic expansion of the physical evolutions kernels around the Sudakov limit, focussing on DIS and the Drell-Yan process. Each order in the expansion is given by a peculiar `jet' or `soft' (depending on the process) {\em physical} Sudakov anomalous dimension.

\noindent The ansatz has been derived from a large--$\beta_0$  dispersive representation, which can be readily extended to the full non-Abelian theory at finite $\beta_0$. Clearly more justifications and checks are still needed. In this respect, the general OPE method introduced in \cite{Akhoury:1998gs,Sotiropoulos:1999hy}
may give the appropriate tool for a systematic derivation.
Another possibility (which shall be the subject of a future investigation) is to perform a check of the ansatz by matching it with existing \cite{Moch:2004pa,Vermaseren:2005qc}  fixed order calculations;  as a by-product one would get (if the ansatz turns out to be correct) the perturbative expansion of the relevant `jet' and `soft' physical  anomalous dimensions which occur as coefficients in these expansions.

\noindent I have shown that the connection of the ansatz with the dispersive representation requires a non-trivial, and process-dependent, choice of expansion parameter in the Sudakov limit. While the adequate parameter $r$ in the DIS case is standard, and refers to the final state jet mass scale, the meaning
of the corresponding parameter $r_{\DY}$  in the Drell-Yan case remains to be understood: it does {\em not} correspond (as could be naively expected) to the total energy $E_{\DY}$ radiated in the final state. 

\noindent It was  found that, although the {\em leading} jet (${\cal J}(W^2)$) and soft (${\cal S}(W_{\DY}^2)$) momentum space physical anomalous dimensions do not \cite{Gardi:2007ma} have renormalons (at least at large--$\beta_0$),  renormalons do start to appear in the subleading jet (${\cal J}_0(W^2)$) and soft (${\cal S}_0(W_{\DY}^2)$)  physical anomalous dimensions.
 
\noindent Given the multiplicity of emerging physical Sudakov anomalous dimensions, it
 would be interesting to find out whether there exists any  relationship between them.
In particular, one might wonder whether any exact, or approximate, universality
property holds among the various `Sudakov effective charges'  $a_{{\cal J},\,{\cal
J}_i}^{\Mink}$ and $a_{{\cal S},\,{\cal S}_i}^{\Mink}$.
Actually, from the observation that the sum ${\cal J}(W^2)+{\cal J}_0(W^2)$ is a total
derivative at large--$\beta_0$ (which follows from the finitness of ${\cal F}_{{\cal J}}(\xi)+{\cal
F}_{{\cal J}_0}(\xi)$ at
$\xi=0$, see (\ref{Fr-exp})),  and making the assumption that this property still holds at {\em finite} $\beta_0$, one can already readily deduce in the DIS case that
$a_{{\cal J}}^{\Mink}(\mu^2)$ and
$a_{{\cal J}_0}^{\Mink}(\mu^2)$ have identical expansions up to NLO, and therefore
\cite{Gardi:2007ma} coincide with the cusp anomalous dimension up to this order.

\noindent The transition to an $1/N$ expansion in moment space has been shown to be straightforward, and the connection between the momentum and moment space jet (or soft) scales has been clarified. From the expansions in moment space, one can easily obtain  the moment space dispersive representations, following the method in \cite{Gardi:2007ma}, by substituting the dispersive representations of the physical jet and soft physical anomalous dimensions in the moment space expansion.

\noindent The longitudinal structure function $F_L(x,Q^2)$, which has not been addressed in this paper, needs a special treatment (to be reported elsewhere). Indeed, it appears the dispersive approach does not work in a straightforward way in this case, essentially because at large--$\beta_0$ the longitudinal physical evolution kernel $K_L(x,Q^2)$ does not coincide any more with the derivative $dF_L(x,Q^2)/dQ^2$. This is due to the fact that the longitudinal coefficient function is ${\cal O}(\alpha_s)$, rather then ${\cal O}(\alpha_s^0)$ as in the case of $F_2$ (which could be replaced at large--$\beta_0$ by its leading term $\delta(1-z)$ on the right hand side of Eq. (\ref{F_2_evolution})). In particular, $K_L(x,Q^2)$ is {\em not} suppressed for $x\rightarrow 1$, contrary to $F_L(x,Q^2)$ itself. Preliminary investigation nevertheless seems to indicate that an ansatz of the form of Eq. (\ref{r-expansion-DIS}) might still be valid for $K_L(x,Q^2)$ (with a different coefficient of the $\delta(1-x)$ term), although the connection with the dispersive approach is lost (except eventually at large--$\beta_0$, where the leading Sudakov physical anomalous dimension ${\cal J}_{L}\left(W^2\right)$ could be a total derivative).

\noindent Finally,  although I have studied here the simplest examples where only one scale (`jet' or `soft') is present, the ansatz can be readily extended to other cases  \cite{Gardi:2007ma} where both of these scales do occur simultaneously.

\vspace{0.5cm}

\noindent\textbf{Acknowledgements}

\vspace{0.3cm}

\noindent I wish to thank Einan Gardi for a number of past discussions on the dispersive approach.

\appendix

\section{$N\rightarrow\infty$ expansion in moment space (Deep Inelastic Scattering)\label{sec:largeN-DIS}}
The $N\rightarrow\infty$ expansion of the moment space physical evolution kernel  Eq.~(\ref{DIS_kernel_def}) can be straightforwardly obtained by taking the moments of  the momentum space expansion Eq.~(\ref{r-expansion-DIS}) ($r=\frac{1-x}{x}$, $W^2=r\,Q^2$):

\begin{equation}
\label{N-expansion-DIS}
\widetilde{K}(N,Q^2)=\int_0^1dx\,x^{N-1}\,\frac{1}{r}\,{\cal J}\left(W^2\right)\,+\,\frac{d\ln \left({\cal F}(Q^2)\right)^2}{d\ln Q^2}\,+\int_0^1dx\,x^{N-1}\,{\cal J}_{0}\left(W^2\right)\,+...
\end{equation}
where the argument $W^2$ of the `jet' anomalous dimensions, being $x$-dependent, has to be integrated over.
The leading term in this expansion is infrared divergent at $x=1$ ($r=0$), but, as we shall see below, this divergence is regularized by the virtual contribution.
Considering first the subleading terms, which are infrared finite, we have identically ($i\geq 0$):

\begin{equation}
\label{Ji-moments}
\int_0^1dx\,x^{N-1}\,r^i\,{\cal J}_{i}\left(W^2\right)=\frac{1}{N^{1+i}}\int_0^\infty dt\,\left(\frac{1}{1+\frac{t}{N}}\right)^{N+1}\,t^i\,{\cal J}_{i}\left(t\,\frac{Q^2}{N}\right)\,
\end{equation}
where I made the change of variable $t=N\,r$. Then, using the $N\rightarrow\infty$ expansion (with $p=1$)
\begin{equation}
\label{exp}
\left(\frac{1}{1+\frac{t}{N}}\right)^{N+p}={\rm e}^{-t}+\frac{1}{N}\,{\rm e}^{-t}\,\frac{t}{2}(t-2\,p)+{\cal O}(1/N^2)\ ,
\end{equation}
one gets, letting $N\rightarrow\infty$ with the moment space `jet' scale $\widetilde{W}^2\equiv Q^2/N$ {\em fixed} inside the integral on the r.h.s. of (\ref{Ji-moments}):
\begin{equation}
\label{Ji-moments-as}
\int_0^1dx\,x^{N-1}\,r^i\,{\cal J}_{i}\left(W^2\right)=\frac{1}{N^{1+i}}\int_0^\infty dt\,{\rm e}^{-t}\,t^i\,{\cal J}_{i}\left(t\,\widetilde{W}^2\right)+{\cal O}(1/N^{2+i})\ .
\end{equation}
One should note that the integral on the r.h.s. of (\ref{Ji-moments-as}) contains constant terms, as well as logarithms of $N$. Moreover, the integral being convergent, the constant terms determine the logarithmically divergent terms, given the beta function coefficients: see the argument following Eq. 4 in \cite{Grunberg:2006gd}--which  implies that the knowledge of the logarithmic terms at a given subleading order of the $1/N$ expansion {\em also fixes the constant terms} at the same order.
In particular, the $N\rightarrow\infty$ expansion  of the next to leading real emission term in Eq.~(\ref{N-expansion-DIS}) is:
\begin{equation}
\label{J0-moments-as}
\int_0^1dx\,x^{N-1}\,{\cal J}_{0}\left(W^2\right)=\frac{1}{N}\int_0^\infty dt\,{\rm e}^{-t}\,{\cal J}_{0}\left(t\,\widetilde{W}^2\right)+\,{\cal O}(1/N^2)\ .
\end{equation}

Considering now the leading, infrared divergent term in Eq.~(\ref{N-expansion-DIS}), the same procedure yields the analogue of  Eq.~(\ref{Ji-moments}):
\begin{equation}
\label{J-moments}
\int_0^1dx\,x^{N-1}\,\frac{1}{r}\,{\cal J}\left(W^2\right)=\int_0^\infty dt\,\left(\frac{1}{1+\frac{t}{N}}\right)^{N+1}\,\frac{1}{t}\,{\cal J}\left(t\,\frac{Q^2}{N}\right)\ ,
\end{equation}
and, taking the $N\rightarrow\infty$ limit with $Q^2/N$ fixed inside the integral on the r.h.s., the analogue of Eq.~(\ref{Ji-moments-as}):
\begin{equation}
\label{J-moments-as}
\int_0^1dx\,x^{N-1}\,\frac{1}{r}\,{\cal J}\left(W^2\right)=\int_0^\infty dt\,{\rm e}^{-t}\,\frac{1}{t}\,{\cal J}\left(t\,\widetilde{W}^2\right)+{\cal O}(1/N)\ .
\end{equation}
The infrared divergence in Eq.~(\ref{J-moments-as}) can be regularized as in Eq.~(\ref{r-expansion-reg-DIS}) by subtracting from the r.h.s. of (\ref{J-moments-as}) the IR divergent piece $\int_0^{Q^2}\frac{d\mu^2}{\mu^2}{\cal J}(\mu^2)$, and merging it together with the virtual quark form factor contribution $\frac{d\ln \left({\cal F}(Q^2)\right)^2}{d\ln Q^2}$ in Eq.~(\ref{N-expansion-DIS}) (where the IR divergences cancell, as we have seen in section~(\ref{sec:expansion-DIS})). Since $\int_0^{Q^2}\frac{d\mu^2}{\mu^2}{\cal J}(\mu^2)=\int_0^N dt\,\frac{1}{t}\,{\cal J}\left(t\frac{Q^2}{N}\right)$ (where I set $t=N\mu^2/Q^2$), one obtains  the $N\rightarrow\infty$ expansion  of the (regularized) leading  term in Eq.~(\ref{N-expansion-DIS}):
\begin{align}
\label{J-reg-moments-as}
\begin{split}
\int_0^1dx\,x^{N-1}\,\left[\frac{1}{r}\,{\cal J}\left(W^2\right)\right]_{+} =&\left[\int_0^\infty dt\,{\rm e}^{-t}\,\frac{1}{t}\,{\cal J}\left(t\,\widetilde{W}^2\right)-\int_0^N dt\,\frac{1}{t}\,{\cal J}\left(t\,\widetilde{W}^2\right)\right] \\
&+\frac{1}{N}\int_0^\infty dt\,{\rm e}^{-t}\,\frac{1}{2}(t-2)\,{\cal J}\left(t\,\widetilde{W}^2\right)+\,{\cal O}(1/N^2)\ .
\end{split}
\end{align}
Combining (\ref{J0-moments-as}) with (\ref{J-reg-moments-as}) one thus gets the large $N$ expansion of the first two leading real emission terms in Eq.~(\ref{N-expansion-DIS}):
\begin{align}
\label{J+J0-moments-as}
\begin{split}
\int_0^1dx\,x^{N-1}\,\left\{\left[\frac{1}{r}\,{\cal J}\left(W^2\right)\right]_{+}+\,{\cal J}_{0}\left(W^2\right)\right\} =&\left[\int_0^\infty dt\,{\rm e}^{-t}\,\frac{1}{t}\,{\cal J}\left(t\,\widetilde{W}^2\right)-\int_0^N dt\,\frac{1}{t}\,{\cal J}\left(t\,\widetilde{W}^2\right)\right] \\
+&\frac{1}{N}\int_0^\infty dt\,{\rm e}^{-t}\,\left({\cal J}_0\left(t\,\widetilde{W}^2\right)\,+\frac{1}{2}(t-2)\,{\cal J}\left(t\,\widetilde{W}^2\right)\right)\\
+&\,\,\,\,\,\,{\cal O}(1/N^2)\ ,
\end{split}
\end{align}
to which\footnote{The first term on the r.h.s of (\ref{J+J0-moments-as}) can also be written, up to exponentially small corrections at large $N$, as $\int_0^N dt\,({\rm e}^{-t}-1)\,\frac{1}{t}\,{\cal J}\left(t\,\widetilde{W}^2\right)$.} one has to add the (N-independent) regularized virtual contribution $\Big[\frac{d\ln \left({\cal F}(Q^2)\right)^2}{d\ln Q^2}\,+\,\int_0^{Q^2}\frac{d\mu^2}{\mu^2}{\cal J}(\mu^2)\Big]$.

I also note that taking the moments of the  leading term in the first line of Eq.~(\ref{K_x}) one gets (setting $t=N(1-x)$):
\begin{align}
\label{J-moments-as-stan}
\begin{split}
\int_0^1dx\,x^{N-1}\,\frac{1}{1-x}\,{\cal J}\left((1-x)Q^2\right) =&\int_0^N dt\,\left(1-\frac{t}{N}\right)^{N-1}\,\frac{1}{t}\,{\cal J}\left(t\,\frac{Q^2}{N}\right)\\
=&\int_0^\infty dt\,{\rm e}^{-t}\,\frac{1}{t}\,{\cal J}\left(t\,\widetilde{W}^2\right)+{\cal O}(1/N)\ ,
\end{split}
\end{align}
which, compared to (\ref{J-moments-as}), checks the equivalence of the leading terms in (\ref{K_x}) and (\ref{r-expansion-reg-DIS}).

\section{$N\rightarrow\infty$ expansion in moment space (Drell-Yan)\label{sec:largeN-DY}}
The $N\rightarrow\infty$ expansion of the moment space physical evolution kernel  Eq.~(\ref{dG_N}) can similarly be straightforwardly obtained by taking the moments of  the momentum space expansion Eq.~(\ref{r-tilde-expansion-DY}) ($r_{\DY}= 2\,\frac{1-\tau}{\tau+\sqrt{\tau}}$, $W_{\DY}^2=r_{\DY}^2\,Q^2$):

\begin{equation}
\label{N-expansion-DY}
\widetilde{K}_{\DY}(N,Q^2)=\int_0^1d\tau\,\tau^{N-1}\,\frac{2}{r_{\DY}}\,{\cal S}\left(W_{\DY}^2\right)\,+\,\frac{d\ln \left|\left({\cal F}(-Q^2)\right)\right|^2}{d\ln Q^2}\,\delta(1-\tau)\,+\int_0^1d\tau\,\tau^{N-1}\,{\cal S}_{0}\left(W_{\DY}^2\right)\,+...
\end{equation}
The leading term in this expansion is infrared divergent at $\tau=1$ ($r_{\DY}=0$), but, as we shall see below, this divergence is regularized by the virtual contribution.
Considering first the subleading terms, which are infrared finite, we have identically ($i\geq 0$):

\begin{equation}
\label{Si-moments}
\int_0^1d\tau\,\tau^{N-1}\,r_{\DY}^i\,{\cal S}_{i}\left(W_{\DY}^2\right)=\frac{1}{N^{1+i}}\int_0^\infty du\,\left(\frac{1}{1+\frac{u}{2\,N}}\right)^{2\,N+1}\,u^i\,{\cal S}_{i}\left(u^2\frac{Q^2}{N^2}\right)\,
\end{equation}
where I made the change of variable $u=N\,r_{\DY}$, which, remarkably, leads to an integral of a form quite similar to the one on the r.h.s. of (\ref{Ji-moments}). Then, using the $N\rightarrow\infty$ expansion (\ref{exp})
one gets, letting $N\rightarrow\infty$ with the moment space `soft' scale $Q^2/N^2\equiv\widetilde{W}_{\DY}^2$ {\em fixed} inside the integral on the r.h.s. of (\ref{Si-moments}):
\begin{equation}
\label{Si-moments-as}
\int_0^1d\tau\,\tau^{N-1}\,r_{\DY}^i\,{\cal S}_{i}\left(W_{\DY}^2\right)=\frac{1}{N^{1+i}}\int_0^\infty du\,{\rm e}^{-u}\,u^i\,{\cal S}_{i}\left(u^2\,\widetilde{W}_{\DY}^2\right)+{\cal O}(1/N^{2+i})\ .
\end{equation}
Thus the $N\rightarrow\infty$ expansion  of the next to leading real emission term in Eq.~(\ref{N-expansion-DY}) is:
\begin{equation}
\label{S0-moments-as}
\int_0^1d\tau\,\tau^{N-1}\,{\cal S}_{0}\left(W_{\DY}^2\right)=\frac{1}{N}\int_0^\infty du\,{\rm e}^{-u}\,{\cal S}_{0}\left(u^2\,\widetilde{W}_{\DY}^2\right)+{\cal O}(1/N^{2})\ .
\end{equation}

Considering now the leading, infrared divergent term in Eq.~(\ref{N-expansion-DY}), the same procedure yields the analogue of  Eq.~(\ref{Si-moments}):
\begin{equation}
\label{S-moments}
\int_0^1d\tau\,\tau^{N-1}\,\frac{2}{r_{\DY}}\,{\cal S}\left(W_{\DY}^2\right)=2\int_0^\infty du\,\left(\frac{1}{1+\frac{u}{2\,N}}\right)^{2\,N+1}\,\frac{1}{u}\,{\cal S}\left(u^2\frac{Q^2}{N^2}\right)\ ,
\end{equation}
and, taking the $N\rightarrow\infty$ limit with $Q^2/N^2$ fixed inside the integral on the r.h.s. of (\ref{S-moments}), the analogue of Eq.~(\ref{Si-moments-as}):
\begin{equation}
\label{S-moments-as}
\int_0^1d\tau\,\tau^{N-1}\,\frac{2}{r_{\DY}}\,{\cal S}\left(W_{\DY}^2\right)=2\int_0^\infty du\,{\rm e}^{-u}\,\frac{1}{u}\,{\cal S}\left(u^2\,\widetilde{W}_{\DY}^2\right)+{\cal O}(1/N)\ .
\end{equation}
The infrared divergence in Eq.~(\ref{S-moments-as}) can be regularized as in Eq.~(\ref{r-expansion-reg-DY})  by subtracting from the r.h.s. of (\ref{S-moments-as}) the IR divergent piece $\int_0^{Q^2}\frac{d\mu^2}{\mu^2}{\cal S}(\mu^2)$, and merging it together with the virtual quark form factor contribution $\frac{d\ln \left|\left({\cal F}(-Q^2)\right)\right|^2}{d\ln Q^2}$ in Eq.~(\ref{N-expansion-DY}) (where the IR divergences cancell, as we have seen in section~(\ref{sec:expansion-DY})). Since $\int_0^{Q^2}\frac{d\mu^2}{\mu^2}{\cal S}(\mu^2)=2\int_0^N du\,\frac{1}{u}\,{\cal S}\left(u^2\frac{Q^2}{N^2}\right)$ (where I set $u^2=N^2\mu^2/Q^2$), one obtains  the $N\rightarrow\infty$ expansion  of the (regularized) leading  term in Eq.~(\ref{N-expansion-DY}):
\begin{align}
\label{S-reg-moments-as}
\begin{split}
\int_0^1d\tau\,\tau^{N-1}\,\left[\frac{2}{r_{\DY}}\,{\cal S}\left(W_{\DY}^2\right)\right]_{+} =&\left[2\int_0^\infty du\,{\rm e}^{-u}\,\frac{1}{u}\,{\cal S}\left(u^2\,\widetilde{W}_{\DY}^2\right)-2\int_0^N du\,\frac{1}{u}\,{\cal S}\left(u^2\,\widetilde{W}_{\DY}^2\right)\right] \\
&+\frac{1}{N}\int_0^\infty du\,{\rm e}^{-u}\,\frac{1}{2}(u-2)\,{\cal S}\left(u^2\,\widetilde{W}_{\DY}^2\right)+\,{\cal O}(1/N^2)\ .
\end{split}
\end{align}
Combining (\ref{S0-moments-as}) with (\ref{S-reg-moments-as}) one thus gets the large $N$ expansion of the first two leading real emission terms in Eq.~(\ref{N-expansion-DY}):
\begin{align}
\label{S+S0-moments-as}
\begin{split}
\int_0^1d\tau\,\tau^{N-1}\,\left\{\left[\frac{2}{r_{\DY}}\,{\cal S}\left(W_{\DY}^2\right)\right]_{+}+\,{\cal S}_{0}\left(W_{\DY}^2\right)\right\} =&\left[2\int_0^\infty du\,{\rm e}^{-u}\,\frac{1}{u}\,{\cal S}\left(u^2\,\widetilde{W}_{\DY}^2\right)-2\int_0^N du\,\frac{1}{u}\,{\cal S}\left(u^2\,\widetilde{W}_{\DY}^2\right)\right] \\
+&\frac{1}{N}\int_0^\infty du\,{\rm e}^{-u}\,\left({\cal S}_0\left(u^2\,\widetilde{W}_{\DY}^2\right)\,+\frac{1}{2}(u-2)\,{\cal S}\left(u^2\,\widetilde{W}_{\DY}^2\right)\right)\\
+&\,\,\,\,\,\,{\cal O}(1/N^2)\ ,
\end{split}
\end{align}
to which\footnote{The first term on the r.h.s of (\ref{S+S0-moments-as}) can also be written, up to exponentially small corrections at large $N$, as $2\int_0^N du\,({\rm e}^{-u}-1)\,\frac{1}{u}\,{\cal S}\left(u^2\widetilde{W}_{\DY}^2\right)$.} one has to add the (N-independent) regularized virtual contribution $\Big[\frac{d\ln \left|\left({\cal F}(-Q^2)\right)\right|^2}{d\ln Q^2}\,+\,\int_0^{Q^2}\frac{d\mu^2}{\mu^2}{\cal S}(\mu^2)\Big]$.

I also note that taking the moments of the  leading term in the first line of Eq.~(\ref{K_x_DY}) one gets (setting $u=N(1-\tau)$):
\begin{align}
\label{S-moments-as-stan}
\begin{split}
\int_0^1d\tau\,\tau^{N-1}\,\frac{2}{1-\tau}\,{\cal S}\left((1-\tau)^2\,Q^2\right) =&2\int_0^N du\,\left(1-\frac{u}{N}\right)^{N-1}\,\frac{1}{u}\,{\cal S}\left(u^2\frac{Q^2}{N^2}\right)\\
=&2\int_0^\infty du\,{\rm e}^{-u}\,\frac{1}{u}\,{\cal S}\left(u^2\,\widetilde{W}_{\DY}^2\right)+{\cal O}(1/N)\ ,
\end{split}
\end{align}
which, compared to (\ref{S-moments-as}), checks the equivalence of the leading terms in (\ref{K_x_DY}) and (\ref{r-expansion-reg-DY}).

\end{document}